\documentstyle[12pt,aps,prd,tighten,preprint,eqsecnum]{revtex}
\def\ra{\rightarrow}
\begin{document}
\thispagestyle{empty}
\preprint{\baselineskip 18pt {\vbox{\hbox{SU-4240-581} \hbox{OKHEP-94-10}}}}
\vspace{20pt}
\title{ Heavy Meson Radiative Decays and Light Vector Meson Dominance }
\vspace{25pt}
\author{Pankaj Jain}
\address{Department of Physics and Astronomy, University of Oklahoma, Norman,
OK 73019  }
\date{May,1994}
\author{ Arshad Momen and Joseph Schechter}
\address{ Physics Department, Syracuse University, Syracuse,
New York 13244-1130}
\maketitle
\vspace{35pt}
\begin{abstract}
Electromagnetic interactions are introduced in the effective chiral Lagrangian
for heavy mesons which includes light vector particles. A suitable notion of
vector meson dominance is formulated. The constraints on  the heavy meson
-light vector
and heavy meson-light pseudoscalar coupling constants are obtained
using experimental  $D^* \ra D \, \gamma $
branching ratios. These constraints are compared with values estimated from
semi-leptonic transition amplitudes as well as from extension of the light
meson coupling pattern. Application to the heavy baryon spectrum in the ``bound
state " model is made.
\end{abstract}
\pagebreak
\vfill

\section{Introduction}

Effective Lagrangians combining heavy quark symmetry and chiral invariance
\cite{Wise} provide promising tools for understanding the ``soft" interactions
of the heavy mesons. The apparent dominance of the decays $D \ra {\bar K}^* \,
l \nu
$  over $D \ra {\bar K} \pi l \nu $ \cite{sumati}, as well as general
considerations, suggest the inclusion of the light vector mesons in addition
to the
light pseudoscalars.The total Lagrangian is the sum  of  a ``light" part
describing the three flavors  $u,d,s$ and a ``heavy" part  describing the
``heavy" meson multiplet $H$ and its interaction with the light sector:

\begin{equation}
{\cal L}_{eff} = {\cal L}_{light} + {\cal L}_{heavy} \qquad.
\label{action}
\end{equation}
The relevant light fields belong to the $3\times3$ matrix of pseudoscalars,
$\phi$,
and to the $3\times3$ matrix of vectors, $\rho_\mu$. It is convenient to define
objects which transform simply  under the action of the
chiral group,

\begin{eqnarray}
\xi=\exp{(\frac{i\phi}{F_\pi})},  \qquad U= \xi^2, \nonumber  \\
A_\mu^L = \xi\rho_\mu \xi^\dagger + {i \over {\tilde{g}}}\xi
\partial_\mu\xi^\dagger,\nonumber \\
A_\mu^R = \xi^\dagger\rho_\mu \xi + {i \over {\tilde{g}}}\xi^\dagger
\partial_\mu\xi,
\nonumber \\
F_{\mu\nu} = \partial_\mu\rho_\nu - \partial_\nu\rho_\mu - i{\tilde{g}}
[\rho_\mu,\rho_\nu],
\label{def}
\end{eqnarray}
where $F_\pi \approx$0.132 GeV and ${\tilde{g}} \approx 3.93$ for a typical
fit.

The heavy multiplet field combining  the
heavy pseudoscalar $P'$ and the heavy vector $Q_\mu'$, both moving with a fixed
4-velocity $V_\mu$ is given by :
\begin{equation}
H = {{1-i\gamma_\mu V_\mu}\over 2}(i\gamma_5 P'+ i\gamma_\nu Q'_\nu),
\;\; {\bar H} \equiv \gamma_4 H^\dagger \gamma_4.
\label{1}
\end{equation}
In our convention $H$ has the canonical dimension one.

The light part of the action under consideration has been most recently
discussed in \cite{Big}. Apart from SU(3) and chiral symmetry breaking terms
and terms proportional to the Levi-Civita symbol, it may be written as

\begin{equation}
{\cal L}_{light}= - \frac{1}{4} Tr \, \left( F_{\mu \nu}(\rho)F_{\mu \nu}(\rho)
\right) - \frac{m_v^2(1+k)}{8k} Tr \, \left( A^L_{\mu}A^L_{\mu}
+ A^R_{\mu}A^R_{\mu}\right)
+\frac{m_v^2(1-k)}{4k} Tr \, \left( A^L_{\mu}U A^R_{\mu}U^{\dagger}\right),
\label{light}
\end{equation}
where $m_v\approx0.77\; GeV$ is the light vector mass and $k=
\frac{m_v^2}{(F_{\pi}{\tilde g})^2}$. An alternate ``hidden symmetry"
approach \cite{bando} leads to the identical Lagrangian.

${\cal L}_{heavy} $ has been discussed by several authors
\cite{them,Hvec,ko,kamal}.  Following the notations of
ref. \cite{Hvec}, it is , to leading order in $M$:

\begin{equation}
{{{\cal L}_{heavy}} \over {M}} = iV_\mu Tr \left[ H\left( \partial_\mu -
i\alpha{\tilde{g}}\rho_\mu - i(1-\alpha)v_\mu\right){\bar
H}\right]+id\;Tr\left[
H\gamma_\mu\gamma_5 p_\mu{\bar H}\right] +{ic \over{m_v}}\;Tr \left[ H
\gamma_\mu
\gamma_\nu F_{\mu\nu}(\rho){\bar H}\right],
\label{hea}
\end{equation}
where $M$ is the mass of the heavy meson and
\begin{equation}
v_\mu, p_\mu \equiv {i \over 2} (\xi \partial_\mu \xi^\dagger \pm \xi^\dagger
\partial_\mu \xi).
\end{equation}
 $\alpha,c,d$ are dimensionless
coupling constants for the heavy-light interactions; they are crucial for
discussing the soft dynamics of the heavy mesons as well as other applications.
In particular, we are interested in the dynamics of the heavy baryons in the
bound state model, wherein all three parameter enter in an important way .The
object of this note is to obtain useful restrictions on these three parameters
for the purpose of studying the heavy baryons. A key ingredient in the analysis
will be estimates of the $D^* \rightarrow D \, \gamma$ rates based on a
suitable notion of vector meson dominance for the electromagnetic interactions
of heavy mesons . Analogous restrictions have been discussed
\cite{Amundson,Georgi,Cheng} for the model in which light vectors are not
present . The model where vectors are included was used to calculate the
rates  $D^* \rightarrow D \, \gamma$ \cite{Cola} with $c$ and $d$ obtained
from assumed pole fits to semi-leptonic form factors. Especially for $c$, there
is no detailed experimental confirmation that the form factors have exactly the
pole dependence. Hence, it is desirable to proceed  in a more general way. Here
we shall use a similar model to get estimated bounds on $c$ and $d$. In
addition, we will formulate the model in such a way that vector meson dominance
can be relaxed and deviations calculated.
 At the present stage
it seems most reasonable to work in the leading order in $M$ as well as tree
order in the light fields. Before proceeding we make some preliminary remarks
about the coupling constants.

The coupling constant $d$ in (\ref{hea}) is related to the $D^* \ra D \, \pi$
decay widths as follows,

\begin{eqnarray}
\Gamma( D^{*0} \ra D^0 \, \pi^0 ) = \frac {d^2 p_{00}^3}{12\pi F_{\pi}^2},
\nonumber \\
\Gamma( D^{*+} \ra D^0 \, \pi^+ ) = \frac {d^2 p_{0+}^3}{6\pi F_{\pi}^2},
\nonumber \\
\Gamma( D^{*+} \ra D^+ \, \pi^0 ) = \frac {d^2 p_{+0}^3}{12\pi F_{\pi}^2},
\label{3}
\end{eqnarray}
where $p_{ab}$ is the decay 3-momentum in the parent rest frame for the $D^a \,
\pi^b$ final state. Note that $D^{*0} \ra D^+ \pi^-$ in addition to all of the
$B^* \ra B \, \pi$ decays are energetically forbidden. $D^{*+}_s \ra D^+_s \,
\pi^0$ is energetically allowed but is suppressed \cite{chowise} due to isospin
conservation and has not yet been observed. An experimental bound,
$\Gamma_{total}(D^{*+}) < 131 \,{\rm KeV}$ \cite{ACC}, gives using (\ref{3}),
the restriction
\begin{equation}
| d|< \, 0.70
\label{4}
\end{equation}
If we were to go to higher orders in a $\frac{1}{M}$ expansion , the $d$
appearing in (\ref{3}) and (\ref{4}) should be interpreted as an effective one
\cite{randall} which deviates by a small amount from the $d$ parameter defined
in (\ref{hea}).

The coupling constant $\alpha$ in (\ref{hea}) was introduced in \cite{Hvec} as
a
measure of vector meson dominance in the light-heavy direct interaction. We
will also verify here that $\alpha = 1 $ corresponds to vector meson dominance
for the diagonal matrix elements of the light electromagnetic current
between heavy meson states,$\langle A|J^{light}_\mu|A\rangle$,
where $A=P$ or $Q_\mu$.

 In section II we briefly discuss some general features of heavy meson
radiative decays in a preliminary way using the constituent quark model.
While the vector meson dominance approach yields amplitudes with the same
general structure, and so cannot be considered essentially superior just for
the purpose of calculating these decays , it does express the amplitudes in
terms of the coupling constant $c$ which is of great interest. The point is
made that, even though it is perhaps unmerited by the ``state of the art" (
including experiments ) to go beyond a leading order in $M$ calculation , there
is an unavoidable ``accidentally" important subleading contribution due to the
fact that two different pieces of the electromagnetic current are being probed.

In section III we discuss how to add the electromagnetic interactions to the
effective chiral Lagrangian. The parameter choices which yield vector meson
dominance are noted. In this limit and in the leading order of the chiral
expansion we compute the $D^* \rightarrow D \, \gamma$ amplitudes. Comparing
with the existing experimental data yields somewhat restrictive bounds on the
coupling constants $d$ and $c$ ( $d$ enters the problem because present
experiments give values for the branching ratios but not the total rates ).

In section IV , the allowed region for $c$ and $d$ is compared with two
attempts at determining these parameters. It is found that if they are obtained
by considering the strange quark to be  the heavy one , we get a point slightly
outside the allowed region. Estimates of $d$ and $c$ based on fitting the $D
\rightarrow K^*$ and $D \rightarrow K$ semi-leptonic decays to pole-form are
noted to lie within the allowed region . The constraints are also used to limit
the prediction of the heavy baryon mass spectrum in the soliton ``bound state "
model.

\section{Heavy Vector Meson Radiative Decays}

The decays of the type $D^* \ra D \,\gamma$ are governed by the fundamental
electromagnetic interaction:

\begin{equation}
{\cal L}_{EM} = e J^{EM}_\mu {\cal A}_\mu
\label{e1}
\end{equation}
where $e$ is the proton charge and ${\cal A}_\mu$ is the photon field.
It is important
to note that we need both the light and the heavy pieces in the decomposition :

\begin{eqnarray}
J^{EM}_\mu = J^{light}_\mu + J^{heavy}_\mu, \nonumber \\
J^{light}_\mu = i\left[ \frac{2}{3} {\bar u}\gamma_\mu u - \frac{1}{3}
 \left({\bar d}\gamma_\mu d + {\bar s}\gamma_\mu s \right)\right], \nonumber \\
J^{heavy}_\mu = i{\cal C} {\bar Q}\gamma_\mu Q + \cdots,
\label {e2}
\end{eqnarray}
where ${\cal C}$ is the electric charge of the particular heavy quark under
consideration (e.g. $\frac{2}{3}$ for the $c$ quark ).

For orientation purposes it is useful to consider how the $D^* \ra D \, \gamma
$ decays are computed in the simplest non-relativistic ``constituent" quark
model \cite{rosnerrev}. The interaction Hamiltonian is $-{\bbox{\mu\cdot B}}$
where the magnetic moment operator is
\begin{equation}
{\bbox{\mu}} = e\left( \frac{{\cal C}}{M} {\bbox{S}}^{heavy} +
\frac{\bar q}{m} {\bbox{S}}^{light} \right)  ,
\label{e3}
\end{equation}
while $M$ and $m$ are the heavy and light contituent quark masses respectively.
${\bar q}$ is the charge of the the light anti-quark in the heavy meson. The
two terms in (\ref{e3}) illustrate the decomposition of $J^{EM}_\mu$ in
(\ref{e2}). After sandwiching (\ref{e3}) between vector and pseudoscalar spin
wavefunctions we find that the amplitude for $D^* \ra D \, \gamma$ is
proportional to $(\frac{{\cal C}}{M} - \frac{\bar q}{m})$. This means that the
amplitudes for $D^{*0} \ra D^0 \, \gamma,\qquad D^{*+} \ra D^+ \, \gamma,\qquad
$ and $D^{*+}_s \ra D^+_s \, \gamma$ stand in the ratios :
\begin{equation}
(1+\frac{m}{M}) : (-\frac{1}{2} + \frac{m}{M}) :
(-\frac{m}{2m_s}+\frac{m}{M})
\label{e4}
\end{equation}
where $m_s$ is the  constituent strange quark mass. (For comparison the
corresponding radiative amplitudes for ${\bar B}^{*-} \ra {\bar B}^- \,
\gamma,\qquad {\bar B}^{*0} \ra {\bar B}^0 \, \gamma$
and ${\bar B}^{*0}_s \ra {\bar B}^0_s \, \gamma$ stand in the ratios
$(-2+\frac{m}{M}) : (1 + \frac{m}{M}) :
(\frac{m}{m_s}+\frac{m}{M})$). Now we would like to start working in the
leading $M \ra \infty$ limit. However, (\ref{e4}) indicates that this is a
rather questionable approximation for $ D^* \ra D \, \gamma$ since $m \approx
0.35$ GeV while $M$ is in the 1.5 - 1.8 GeV range. In the case of $D^{*+} \ra
 D^+ \, \gamma$ the piece proportional to $\frac{1}{M}$  is expected to be
almost half
of the leading term and opposite in sign. We note that the structures of
$J_\mu^{light}$ and $J_\mu^{heavy}$ are essentially different so there is no
reason to expect that $J_\mu^{light}$ has large $\frac{1}{M}$ corrections. Thus
we consider the present calculation to be correct to order 1 , but wherein the
``accidentally " important $\frac{1}{M}$ contamination due to the leading
$J_\mu^{heavy}$ has been included . That $\frac{1}{M}$ piece is fixed
\cite{Georgi}
from its relation to the heavy quark number current and corresponds to a term
in the effective Lagrangian:
\begin{equation}
-\frac{1}{2} e {\cal C}{\cal A}_\mu Tr [ \partial_\nu ( H{\bar H} \sigma _{\mu
\nu}) ]  .
\label{e5}
\end {equation}

\section{Vector Meson Dominance}

First, let us review how to add electromagnetic interactions to the light
particle Lagrangian (\ref{light}). More details are given in \cite{joe} and in
section III of \cite{jain} {\footnote{ ${\tilde g}$
is called $g$ in these references}}. The fields $A_\mu^L$ and $A_\mu^R$
introduced in (\ref{def}) are taken to transform under a local chiral
transformation, $U_{L,R} = 1 + E_{L,R}$, as

\begin{equation}
\delta A_\mu^{L,R} = - [ A_\mu^{L,R} , E_{L,R} ] -\frac{i}{\tilde g}
\partial_\mu E_{L,R}.
\label{v1}
\end{equation}

External fields, $B_\mu^{L,R}$ transform as
\begin{equation}
\delta B_\mu^{L,R} = - [ B_\mu^{L,R} , E_{L,R} ] -\frac{i}{h}
\partial_\mu E_{L,R},
\label{v2}
\end{equation}
where $h$ is the external field coupling constant. Then it is clear that
(\ref{light}) will become locally gauge invariant if we make the substitutions
$A_\mu^{L,R} \ra A^{L,R}_\mu - \frac{h}{\tilde g} B_\mu^{L,R}$.
Here we are interested in the
case of electromagnetism, which corresponds to the choice
\begin{equation}
hB_\mu^{L,R} = eQ{\cal A}_\mu, \qquad \qquad Q=diag ( \frac{2}{3},
-\frac{1}{3},-\frac{1}{3}).
\label{v3}
\end{equation}
The resulting electromagnetic interaction piece from (\ref{light}) may be
expanded
out ( see (14) of \cite{joe}) to yield the leading relevant terms
\begin{equation}
e{\cal A}_\mu [ k{\tilde g} F_\pi^2 Tr (Q\rho_\mu) +i (1- \frac{k}{2}) Tr [ Q (
\phi \partial_\mu \phi - \partial_\mu \phi \phi )]]+\cdots \qquad \cdot
\label{v4}
\end{equation}
It is possible to eliminate the photon-vector meson cross term by
rediagonalization but it is more conventional to keep it in this form.
Notice that  the special choice $k=2$ is denoted the
Kawarabayashi-Suzuki-Riazuddin-Fayazuddin (KSRF) relation \cite{ksfr}.
When this holds
the second term in (\ref{v4}) vanishes and the photon couples to the charged
light pseudoscalars via its mixing with the light neutral vector mesons in the
first term. Actually $k$ seems to be 10\% higher than the value required by
the KSRF relation so that this picture is reasonably good but not perfect.

A leading $O(M^0)$ contribution to the $D^* \ra D\, \gamma$ decays arises via
the $c$ term in (\ref{hea})  ( which is locally chiral invariant) giving $D^*
D \rho^0$ and $D^* D \omega^0$ vertices followed by the $\rho^0-\gamma$ and
$\omega^0-\gamma$ mixings in (\ref{v4}). In addition, it is possible to
construct a direct $H{\bar H}\gamma$ locally chiral invariant interaction term
as :
\begin{equation}
\frac{ieM\delta}{2m_v} Tr( H\gamma_\mu \gamma_\nu [ \xi^{\dagger} F_{\mu
\nu}(B_L)\xi + \xi F_{\mu \nu}(B_R)\xi^{\dagger}]{\bar H}), \nonumber
\end{equation}
where we used the facts that under local chiral transformations $\xi {\bar H}
\ra U_L (x)\xi {\bar H}$ and $\xi^\dagger {\bar H} \ra U_R(x)\xi^\dagger
{\bar H}$.
Specializing this  to electromagnetic external fields using (\ref{v3}) gives an
effective term :
\begin{equation}
\frac{ieM\delta}{2m_v}  F_{\mu
\nu}({\cal A})Tr( H\gamma_\mu \gamma_\nu [ \xi^\dagger Q\xi +
\xi Q \xi^{\dagger}]{\bar H}),
\label{v5}
\end {equation}
whose strength is measured by the parameter $\delta$.

Putting the contributions to $D^* \ra D \, \gamma$ from [(\ref{hea}) and
(\ref{v4})] and  from (\ref{v5}) together with the subleading one
from (\ref{e5})
finally yields the width expressions:
\begin{eqnarray}
\Gamma( D^{*0} \ra D^0 \, \gamma ) = \frac {e^2 p_{0\gamma}^3}{12\pi}\left[
\frac{2}{3M} + \frac{8}{3m_v}(\delta+ \frac{c}{\tilde g})\right]^2,
\nonumber \\
\Gamma( D^{*+} \ra D^+ \, \gamma ) = \frac {e^2 p_{+\gamma}^3}{12\pi}\left[
\frac{2}{3M} - \frac{4}{3m_v}(\delta+ \frac{c}{\tilde g})\right]^2,
\nonumber \\
\Gamma( D^{*+}_s \ra D^+_s \, \gamma ) \approx \Gamma( D^{*+} \ra D^+ \, \gamma
),
\label{v6}
\end{eqnarray}
wherein $p_{0\gamma}$ and $p_{+\gamma}$ are the 3-momenta in the $D^{*0}$ and
$D^{*+}$ rest frames respectively. The third approximate equality utilizes the
coincidence that the phase space factors are approximately equal for the two
reactions. We shall not make use of the $D^{*+}_s \ra D^+_s \, \gamma$
reaction here. To compute it more accurately in the present model (even at the
tree level) involves taking into account several SU(3) symmetry breaking terms
discussed in \cite{Big}. Notice that the structure of the amplitudes in
(\ref{v6}) is essentially the same as that in the naive quark model,
(\ref{e4}).
The parameter $M$ in (\ref{v6}), however, is more reasonably considered to be
the heavy meson mass. A natural notion of light vector meson dominance for $
D^* \ra D \,\gamma$ is to set
\begin{equation}
\delta=0
\label{v7}
\end{equation}
which corresponds to the photon interacting with the light electromagnetic
transition moment {\it{only}} through its mixing with the light vectors
in (\ref{v4}). Of course, there is no {\it{ a priori}} reason for the
assumption (\ref{v7}) to be perfect, but usual low energy phenomenology
suggests that it is very sensible.

Now let us compare with experiment. The latest data from CLEO II \cite{butler}
yields {\footnote{We would like to thank S. Stone  for pointing out to us  that
the value quoted in \cite{butler} for $\frac{\Gamma(D^{*+} \ra D^+ \,
\gamma)}{\Gamma_{tot.}}$ is best interpreted as an upper bound of 4\%}}
\begin{eqnarray}
\frac{\Gamma(D^{*0} \ra D^0 \, \gamma)}{\Gamma(D^{*0} \ra D^0 \, \pi^0)}=0.57
\pm 0.14, \nonumber \\
\frac{\Gamma(D^{*+} \ra D^+ \, \gamma)}{\Gamma(D^{*+} \ra D^0 \, \pi^+)} <
0.059.
\label{v8}
\end{eqnarray}

These numbers should be equated to the predictions from (\ref{v6}) and
(\ref{3}). Defining, for temporary convenience,
\begin{equation}
A= \frac{2}{3Md}, \qquad \qquad B=\frac{4}{3m_v d} (\delta + \frac{c}{\tilde
g}),
\label{v9}
\end{equation}
we then get $|A+2B|$ = 3.40 $\pm 0.42 \,{\rm GeV}^{-1}$ as well as $|A-B|
<$ 1.37
${\rm GeV}^{-1}$. Taking $B > 0$ , the allowed region{\footnote{Another region is obtained by reversing the signs of
both $A$ and $B$. However, this choice would disagree with the sign of
(\ref{d3}).} is shown in Fig. 1.
Notice that $B$ is  leading in $M$ while $A$ represents the subleading
contribution. Hence it is natural to consider $B\,> \,A$ as the region of
interest. This correspponds to the allowed region OPQR. (We
shall not consider further here the region PTSQ for which $A > B$ ).
Furthermore the bound (\ref{4}) would require $A$ to lie to the right of the
dashed line (obtained by setting $M$ =1.8 GeV in (\ref{v9}). We see that $B$ is
 restricted to lie between 1 and 1.65. With the vector
meson dominance assumption that $\delta = 0$, this leads to the bounds on the
interesting $\frac{c}{d}$ ratio

\begin{equation}
2.27 < \frac{c}{d} < 3.75.
\label {interest}
\end{equation}
Choosing $M$= 1.8 GeV and comparing with (\ref{4}) shows that 0.29 $< \,d \,
<$ 0.70.

It should be stressed that the constraints on $c$ and $d$ just discussed have
been based on the assumptions
\begin{itemize}
\item {i) light vector meson dominance ,}
\item{ii) leading order in $\frac{1}{M}$ for $J_\mu^{light}$,}
\item{iii) SU(3) invariance,}
\item{iv) leading order in chiral perturbation theory.}
\end{itemize}
( To some extent ii) can be handled by considering $c$ and $d$ to be effective
values rather than the ones defined in (\ref{hea}).) We thus expect that the
constraints can be systematically improved . Nevertheless, it seems to us very
worthwhile to give an idealized analysis which is suited both to the present
experimental accuracy and the accuracy of many present applications.

To end this section we comment on the first, kinetic term in (\ref{hea}), which
contains the ``chiral derivative" ${\cal D}_\mu {\bar H} = [\partial_\mu -i
\alpha {\tilde g}\rho_\mu - i (1-\alpha) v_\mu]{\bar H}$. The presence of
${\cal D}_\mu$ guarantees invariance under {\it{global}} transformations
belonging to the chiral group. When including electromagnetism we should
naturally add the terms $ -ie{\cal A}_\mu ( Q - {\cal C})$ to ${\cal D}_\mu $.
This is not sufficient since both $v_\mu$ and $\rho_\mu$ pick up inhomogenous
pieces under {\it{local}} electromagnetic U(1) transformations. Thus, we
should replace them by the properly covariant quantities
\begin{eqnarray}
{\tilde v}_\mu = v_\mu + e{\cal A}_\mu [\frac{1}{2}(\xi Q \xi^\dagger
+\xi^\dagger Q \xi) - Q],\nonumber \\
{\tilde \rho_\mu} = \rho_\mu - \frac{e}{{\tilde g}}Q{\cal A}_\mu.\nonumber
\end{eqnarray}
The net result is an ``electrified chiral derivative" to be used in
(\ref{hea}):
\begin{eqnarray}
{\cal D}_\mu^{ECD}{\bar H} &=& [ \partial_\mu - ie{\cal A}_\mu(Q-{\cal C}) -
i\alpha{\tilde g}{\tilde \rho}_\mu - i( 1- \alpha) {\tilde v}_\mu ]{\bar
H}\nonumber \\
&=& {\cal D}_\mu {\bar H} + ie {\cal A}_\mu[{\cal C} -
\frac{1}{2}(1-\alpha)(\xi
Q \xi^\dagger + \xi^\dagger Q \xi ) ] {\bar H}.
\label{v13}
\end {eqnarray}
 From this expression, it is evident that the choice $\alpha=1$ corresponds to
no direct photon coupling to the {\it{light}} part of the heavy meson
field ${\bar H}$. The indirect coupling via photon mixing with the light
vectors ensures proper normalization of the electromagnetic form factors of the
heavy meson at zero momentum transfer. Since the first term in (\ref{hea}) is
diagonal,of course, it
does not contribute to the off-diagonal transition matrix elements of interest.
Even though $\alpha=1$ has been seen to be the choice for vector meson
dominance of
the diagonal matrix elements of light electromagnetic currents between heavy
meson states, one still does not know experimentally
 just how good that assumption really is.

\section{Other Estimates and An Application}

We first
consider two other estimates for the light pseudoscalar-heavy meson coupling
constant $d$ and the light vector-heavy meson coupling constant $c$.

One way to get a handle on $c$ and $d$  is to imagine that the flavor SU(3)
invariant expression for the vector-vector-pseudoscalar interaction [ see
(2.18) of \cite{jain}, for example ]:
\begin{equation}
{\cal L}_{VV\phi} = - i g_{VV\phi} \epsilon_{\mu \nu \alpha \beta} Tr (
\partial_\mu \rho_\nu \partial_\alpha \rho_\beta \phi),
\label{d1}
\end{equation}
continues to hold when the strange quark is formally considered to be ``heavy".
Then the $K^+$ field, normally denoted as $\phi_1^3$, would be called
${\bar P}_1$ ,
while the $K^{*+}$ field would be called ${\bar Q}_{1\mu}$ (in the notation of
\cite{Hvec}). If we consider both of the vectors in (\ref{d1}) to be heavy
(with the pseudoscalar  light) the resulting $Q\phi {\bar Q}$
interaction is actually part of the $d$-term in (\ref{hea}) (see (3.20) of
\cite{Hvec}). Similarly the $P\rho {\bar Q}$ piece from (\ref{d1}) is part of
the $c$-term in (\ref{hea}) (see (3.22) of \cite{Hvec} ). These identifications
give the estimates:
\begin{equation}
-g_{VV\phi} = \frac{2d}{F_\pi} = \frac{4c}{m_v}.
\label{d2}
\end{equation}

We immediately see that the ratio:
\begin{equation}
 \frac{c}{d} = \frac{m_v}{2F_\pi}=2.92
\label{d3}
\end{equation}
is compatible with (\ref{interest}). A typical estimate \cite{jain},
$|g_{VV\phi}F_\pi| \approx$ 1.8 yields the additional expectation that $|d|
\approx$ 0.9. The resulting point is denoted $x$ in Fig. 1; it is slightly
outside the allowed region.

A more direct experimental approach to finding  $d$ is based on fitting
\cite{anj}
the $D \ra K$ semi-leptonic transition form factor to $\frac{R}{q^2
+m^2(D^*_s)}$, where $R$=constant. In the present model, $R=
\frac{dM^2F_D}{F_\pi}$, where $M$ is the $D$ mass. This leads to ( see (5.2) of
\cite{us})
\begin{equation}
|d| \approx 0.53,
\label{d4}
\end{equation}
where $F_D \approx$ 0.25 GeV was taken \cite{stone}. A similar approach to
finding $c$ can be based on the study of the $D \ra K^*$ semi-leptonic
transition vector type form factor. There is more theoretical as well as
experimental uncertainty in this case but
one gets \cite{kodama,them,sumati} :
\begin{equation}
|c| \approx 1.6.
\label{d5}
\end{equation}
The point $y$ in Fig. 1 corresponds to choosing both (\ref{d4}) and (\ref{d5});
it clearly lies within the allowed region.

It may be helpful to give some predictions resulting from
the typical parameter choices $d=0.53, c=1.6, \delta =0 $
and $ M=1.8 $ GeV. Then the branching ratios in (3.9) turn out to be
$\Gamma(D^{*0} \ra D^0 \gamma)/ \Gamma(D^{*0} \ra D^0 \pi^0 )=
0.55$ and $ \Gamma(D^{*+} \ra D^+ \gamma)/ \Gamma(D^{*+} \ra D^0
\pi^+)=0.01. $ The total widths are estimated as $
\Gamma_{tot}(D^{*0}) =0.056 $ MeV. and $\Gamma_{tot}(D^{*+})=
0.081 $ MeV. With a heavy mass choice of $ 5.28 $ GeV  the radiative
B widths are estimated as $\Gamma({\bar B}^{*-} \ra {\bar B}^-
 \gamma)= 0.43 $ KeV  and $ \Gamma( {\bar B}^{*0} \ra {\bar B}^0
\gamma) =0.14 $ KeV.

 Now let us discuss the application of this analysis to the computation of the
heavy baryon spectrum in the bound state picture \cite{callan}, which was our
original motivation . In this picture the heavy baryon is treated as a bound
state of the heavy meson and an ordinary light baryon (considered as a Skyrme
soliton). To the first rough approximation the problem reduces to that of relative
motion in a spherical harmonic oscillator potential \cite{Wisegroup,us,chow}
\begin{equation}
V(r) = V_0 + \frac{1}{2} \kappa r^2,
\label{spher}
\end{equation}
where $V_0$ and $\kappa$ are numbers which are computable in terms of the
coupling constants $d,c$ and $\alpha$ as well as the light baryon Skyrme
``profiles". A recent analysis \cite{Anand} has shown that the above quadratic
approximation is not as good for the charmed baryons as it is for the bottom
baryons. Unfortunately, most of the data pertains to the charmed
baryons . The energy levels which follow from (\ref{spher}) are :

\begin{equation}
E_n =V_0 + {\sqrt \frac{\kappa}{\mu}} (n+\frac{3}{2})
\label{levels}
\end{equation}
where $\mu$ is the reduced mass ( which we take as 0.633 GeV for the charmed
baryon case ). The ground state is labeled by $n=0$ . We will compare
(\ref{levels}) with the two pieces of experimental data :

\begin{eqnarray}
E_0 = m(\Lambda_c)- m(N) - m(D)=-0.63{\rm GeV}, \nonumber  \\
E_1 - E_0 = m(\Lambda'_c)-m(\Lambda_c) = 0.31 {\rm GeV},
\label{energy}
\end{eqnarray}
where $m(\Lambda'_c)$ is taken from the experimental evidence of
\cite{refs}.
Now the predictions of the bound state model are given in (5.1) of \cite{us}:

\begin{eqnarray}
V_0 =  -1.19d -0.39c -0.26 \alpha\;{\rm GeV}, \nonumber \\
\kappa= 0.27d+0.12c+0.04\alpha \;{\rm GeV}^3,
\label{para}
\end{eqnarray}
wherein we have reversed the sign of the $\alpha$ term as discussed in
\cite{Anand}. The $d,c$ and $\alpha$ terms correspond to the contributions of
 $\pi$ mesons, $\rho$
mesons and $\omega$ mesons , respectively .

First let us consider the situation when only the $d$ term is present . ( This
corresponds to initial treatments in which the light vector mesons are
neglected \cite{Wisegroup}). For the typical choice $d$= 0.53 as in (\ref{d4})
 we have
$E_0$ = 0.09 GeV and $E_1- E_0$ = 0.48 GeV. Thus the ground state baryon is not
even bound in this case. To improve this situation we want $d$ to be as large
as possible ( going to the left in Fig. 1 according to (\ref{v9})). The
experimental bound (\ref{4}) sets this value as $d$ = 0.7. In this case $E_0$=
-0.02 GeV and $E_1-E_0$=0.54 GeV. Here the ground state is just barely bound .
One might expect \cite{Anand} the binding strength to be increased by perhaps
0.1 GeV when going beyond the quadratic approximation but even so the
prediction is very unsatisfactory when compared to (\ref{energy}).

A very large improvement is obtained by including the light vector mesons.
First it is necessary to specify the parameter $\alpha$ , on which we have no
experimental information. However the fact that light vector meson dominance
has been seen to be reasonable for the $D^* \rightarrow D \gamma$
decays suggests
that the vector meson dominance choice , $\alpha$=1 is also reasonable. Let us
fix $\alpha$=1 , $d$= 0.53 and allow $c$ to vary near its allowed range $2.56
< \frac{c}{d}  < 3.61 $ ( as read off from Fig. 1). Then we find from
(\ref{spher})
and (\ref{para}) that $(E_0, E_1-E_0)$= (-0.31,0.74) GeV,(-0.36,0.76) GeV and
(-0.41,0.80) GeV for $\frac{c}{d}$ = 2.5, 3.0 and 3.5 respectively . It is
clear that the binding energy is now in the right ballpark, but somewhat too
small. The excitation energy $E_1-E_0$ is too large , but it has been observed
\cite {Anand} that this is expected to get significantly improved when we go
beyond the quadratic approximation (\ref{levels}).

We have discussed the introduction of the electromagnetic interaction in the
framework of the effective chiral Lagrangian for heavy mesons which
{\it{includes}} light vector mesons. A suitable notion of light vector
meson dominance was formulated. Application was made to the radiative decays of
the $D^*$ mesons with the goal of determining the light heavy coupling
constants $c$ and $d$. It was found that the acceptable range of values, on
the assumption of vector meson dominance, was compatible with information
extracted from semi-leptonic $D$ decays. The structure of the radiative
amplitudes had the same form as in the simple quark model. Apart from the
necessity to include the $\frac{1}{M}$ piece describing the heavy part of the
electromagnetic current ( which is ``accidentally" enhanced for $D^*$ radiative
decays) we worked to leading order $M^0$. Furthermore, higher derivatives,
loops and SU(3) symmetry breaking were neglected. Together with more accurate
measurements of $\Gamma( D^{*+} \ra D^+\, \gamma)$, these additional
corrections may, in the future, greatly clarify the situation. Finally the
constraints on $c$ and $d$ were used for discussing the predicted heavy baryon
spectrum in the ``bound state model".

\section{Acknowledgement}

We would like to thank G.C. Moneti, N. Horwitz, S. Stone and A. Subbaraman
for helpful discussions. This work was
supported in part  by the U.S. DOE contract
nos. DE-FG-02-85ER40231 and  DE-FG05-91ER-40636.

\begin{figure}
\caption{Physically allowed region for $A$ and $B$.}
\end{figure}

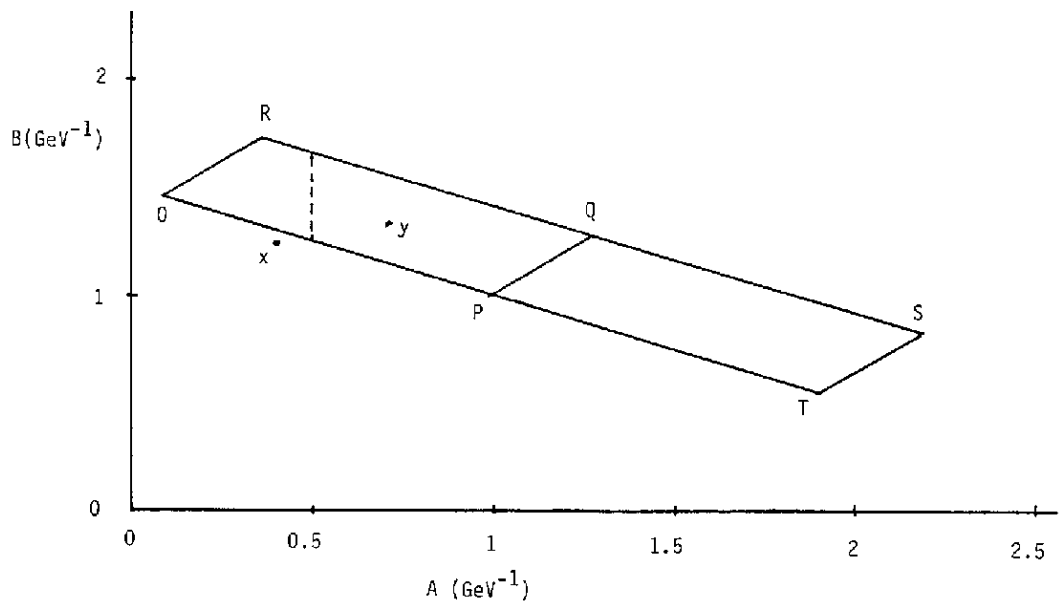
\end{document}